\documentclass[conference]{IEEEtran}
\usepackage{cite}
\usepackage{amsmath,amssymb,amsfonts}
\usepackage{algorithmic}
\usepackage{graphicx}
\usepackage{textcomp}
\usepackage{xcolor}
\usepackage{booktabs}
\usepackage{multirow}
\usepackage{array}
\usepackage{comment}
\usepackage{subcaption}
\usepackage{xurl}
\usepackage{subcaption}
\usepackage{listings}
\usepackage{makecell}

\graphicspath{{Figures/}}
\def\BibTeX{{\rm B\kern-.05em{\sc i\kern-.025em b}\kern-.08em
    T\kern-.1667em\lower.7ex\hbox{E}\kern-.125emX}}
\begin{document}

%\title{Conference Paper Title*\\
%{\footnotesize \textsuperscript{*}Note: Sub-titles are not captured for https://ieeexplore.ieee.org  and
\title{Hype Meets Reality: Large Language Models as Mutators in Search-based Automated Program Repair of Simulink-Stateflow Models}

\author{
\IEEEauthorblockN{Ayesha Irshad}
\IEEEauthorblockA{
\textit{Mondragon University}\\
Mondragon, Spain\\
airshad@mondragon.edu}
\and
\IEEEauthorblockN{Pablo Valle}
\IEEEauthorblockA{
\textit{Mondragon University}\\
Mondragon, Spain\\
pvalle@mondragon.edu}
\and
\IEEEauthorblockN{Jon Ayerdi}
\IEEEauthorblockA{
\textit{Mondragon University}\\
Mondragon, Spain\\
jayerdi@mondragon.edu}
\and
\IEEEauthorblockN{Aitor Arrieta}
\IEEEauthorblockA{
\textit{Mondragon University}\\
Mondragon, Spain\\
aarrieta@mondragon.edu}
}

\maketitle

\begin{abstract}
Search-based Automated Program Repair (APR) techniques rely on carefully designed mutation operators to explore the space of candidate fixes. Recent advances in Large Language Models (LLMs) suggest that generative models could replace such operators by dynamically proposing repairs. In this paper, we investigate this hypothesis in the context of Cyber-Physical Systems (CPSs) modeled in Simulink/Stateflow. We extend the state-of-the-art FlowRepair approach by replacing a subset of its mutation operators with LLM-generated mutations, enabling more flexible and expressive patch generation. We evaluate the approach on a benchmark of 19 real-world faulty Stateflow models across four CPS domains, using the same experimental setup as FlowRepair for controlled comparison under the same wall-clock budget. Contrary to expectations, in this controlled evaluation, the LLM-based mutation substantially degrades repair performance under the FlowRepair experimental setup. Across the tested LLM variants, the LLM-based repair produced plausible patches for 4–6 models and valid patches for 4 models, compared to 18 and 16, respectively, with the original approach. Our analysis reveals that, in this integration, LLMs struggle with precise symbolic edits, lack behavioral feedback, and generate a noisy search space that hinders effective exploration. Rather than showing a general limitation of LLMs for APR, these findings highlight fundamental limitations of naively integrating LLMs into search-based APR and motivate hybrid approaches that combine structured mutation with generative guidance.
\end{abstract}


\section{Introduction}

Cyber-Physical Systems (CPSs) integrate computational elements with physical processes~\cite{derler2011modeling}. Due to their reliance on complex and expensive hardware, their software is typically validated through simulation-based testing during early development stages~\cite{abdessalem2018testing,matinnejad2018test,arrieta2019search}. MATLAB/Simulink has become the de facto platform for CPS modeling and simulation~\cite{matinnejad2018test}, supporting features such as automated test generation and standards-compliant code generation. CPS control architectures commonly separate high-level decision-making from low-level control~\cite{mandrioli2023stress}, with high-level logic often implemented using Stateflow statecharts within Simulink. Although substantial research has addressed automated test generation for Simulink models~\cite{matinnejad2018test,arrieta2017employing,menghi2020approximation,menghi2019generating}, the repair of detected faults remains largely manual, motivating the need for Automated Program Repair (APR) techniques for Simulink models.

A recent study proposed FlowRepair~\cite{arrieta2026flowrepair}, a search-based APR technique for Stateflow models. In addition to novel repair objectives targeting CPSs, FlowRepair~\cite{arrieta2026flowrepair} integrates global and local search and adopts novel repair mutations for Stateflow models. Like many other search-based APR repair tools, FlowRepair~\cite{arrieta2026flowrepair} has one core limitation: The repair capability is limited by the implemented mutation operators. On the one hand, if the number of mutation operators is limited, the tool may not have the ability to repair many types of bugs. On the other hand, if the number of mutation operators is high, the search space may become exceedingly large, making the APR technique not scalable, especially in the context of 
systems with computationally expensive executions.
%these systems where computing the fitness requires executing computationally expensive simulations.

To deal with this problem, in this paper we propose using Large Language Models (LLMs) as mutation operators. We replace key FlowRepair mutation operators with an LLM-based mutator that are able to generate repair candidates directly from the faulty Stateflow model. Instead of relying on a fixed set of predefined mutation templates, the LLM dynamically proposes modifications. This allows the repair process to explore a broader and more flexible search space while avoiding the need to manually design a large number of mutation operators. The generated candidates are then evaluated within the existing search-based repair loop, where FlowRepair's fitness functions assess whether the modified model is closer to resolving the detected bug. 

We evaluate our approach using the same experimental setup as the original FlowRepair paper~\cite{arrieta2026flowrepair} to ensure a controlled comparison under the same wall-clock budget between the traditional mutation-based repair strategy and the proposed LLM-based mutation generation. Surprisingly, the results are substantially worse, indicating that the LLM-based approach does not outperform the original search-based APR technique. In fact, replacing handcrafted mutation operators with LLM-generated mutations substantially reduced repair effectiveness in the observed results in most cases under this experimental setup. While the LLM is able to generate plausible repairs for a small subset of faults, the overall repairing ability and consistency remain substantially lower than those achieved by FlowRepair. These findings highlight important limitations of using LLMs as mutation operators in simulation-based CPS repair contexts and suggest that naively replacing rule-based operators with generative models may degrade repair performance rather than improve it.

% In Section~\ref{sec:background}, we present the general background of FlowRepair, on top of which our approach is developed. In Section~\ref{sec:LLMasmutators} we explain how we substitute those mutation operators with generic LLMs. We assess our approach in Section~\ref{sec:evaluation}, followed by a discussion of why negative results occur in Section~\ref{sec:disc}.In Section~\ref{sec:relatedwork} we position our study against the state-of-the-art. We list the threats to validity in Section~\ref{sec:threats},  and conclude our paper in Section~\ref{sec:conclusion}

\section{Background on How FlowRepair Works}
\label{sec:background}

FlowRepair is a search-based APR technique for repairing faults in
Simulink-Stateflow models of CPSs~\cite{arrieta2026flowrepair}. Unlike traditional APR techniques, which rely on
large, fast-executing test suites, CPS validation requires expensive
simulations of both the controller and the physical environment. This
imposes strict constraints on test execution and available failure
information. In CPS testing, a single simulation may require a complete
system-level execution, where the software controller and physical
plant interact over time. Therefore, repair methods that depend only on
counting passing and failing tests provide limited guidance, especially
when only one failure-inducing test is available, as is common in
falsification-based testing of CPSs\cite{arrieta2026flowrepair}.

To address this, FlowRepair first performs fault localization using
Spectrum-Based Fault Localization (SBFL). The Stateflow model is
instrumented so that the execution of states and transitions can be
recorded for each test case. The available passing and failing tests are
then executed, and the collected execution traces are used to compute
a suspiciousness score for each Stateflow component. FlowRepair uses
the Tarantula metric for this purpose~\cite{arrieta2026flowrepair}. Components that are executed more frequently
by failing tests and less frequently by passing tests receive higher
suspiciousness scores. These scores are then used to rank the states and
transitions according to their likelihood of being faulty. This ranking
does not directly identify the fault, but it guides the repair algorithm
towards the most suspicious parts of the model and reduces unnecessary
mutations on less relevant components~\cite{arrieta2026flowrepair}.

After fault localization, FlowRepair starts the patch generation
process. The repair algorithm receives the faulty Simulink model and
the suspiciousness ranking as input. During the repair process,
FlowRepair maintains two archives. The first archive stores plausible
patches, which are candidate repairs that pass the available test suite.
The second archive stores partial patches, which are modified models
that do not fully pass the failing test but improve the model behaviour
according to the repair objectives. A partial patch is therefore treated
as a useful intermediate repair because it moves the search closer to a
complete fix without necessarily satisfying all requirements yet~\cite{arrieta2026flowrepair}.

Patch generation is guided by a hybrid search strategy that combines
global and local search. The global search is responsible for exploring
different candidate mutations across the suspicious states and
transitions. To select the component to mutate, FlowRepair uses the
suspiciousness ranking through a roulette-wheel selection strategy.
This means that components with higher suspiciousness scores are more
likely to be selected, while components with lower scores can still be
selected with smaller probability. This allows the search to focus on
likely faulty areas without completely ignoring the rest of the model
\cite{arrieta2026flowrepair}. Once a component is selected, FlowRepair applies one compatible
mutation operator and executes the resulting model using
simulation-based testing.

The test execution returns a verdict for the mutated model. If the
mutated model passes the available test suite, FlowRepair stores it in
the plausible patch archive. If the model still fails, FlowRepair checks
whether the verdict has improved compared with the previous model.
A verdict is considered improved when at least one repair objective is
improved and none of the other objectives becomes worse. In that case,
the mutated model is added to the partial patch archive, and the local
search process is triggered~\cite{arrieta2026flowrepair}. This archive-based mechanism is useful
because it preserves promising intermediate solutions instead of
discarding them immediately. Since some partial patches may later lead
to valid repairs, FlowRepair keeps them available for further mutation.

The local search is used to refine promising partial patches. When the
global search finds a mutation that improves the model behaviour,
FlowRepair assumes that the mutated component may be close to the
actual fault. Therefore, the local search focuses on the same Stateflow
component, such as the same state or transition, rather than exploring
the whole model again. This allows the algorithm to exploit a promising
repair direction while reducing the number of expensive simulations.
The local search also considers the mutation operator that produced
the improvement. In some cases, applying the same type of mutation
again may lead to a complete repair, while in other cases another
operator may be needed to modify a related part of the same component
\cite{arrieta2026flowrepair}. To avoid over-exploiting one area or getting trapped in a local
optimum, the number of local attempts is limited.

The search is driven by CPS-specific repair objectives that exploit
temporal execution information. Instead of relying only on the number
of passing and failing tests, FlowRepair evaluates how the candidate
patch changes the failure behavior over time. The first objective is to
reduce the duration for which the failure remains active. A patch that
shortens the failure interval is considered closer to a correct repair.
The second objective is to reduce the severity of the failure, for example
by decreasing the distance between the faulty output signal and the
required threshold. The third objective is to delay the time at which
the failure is triggered. This is particularly useful in CPSs where a
system may not recover once a failure occurs, such as when a controller
drives the system into an unsafe state~\cite{arrieta2026flowrepair}. These objectives provide
more informative guidance than a simple pass/fail verdict and allow
FlowRepair to optimize candidate patches even when only one failing
test case is available.

FlowRepair uses 15 mutation operators tailored to Stateflow models
(explained in Table~\ref{table:MutOperators}). These operators modify states, transitions, and
expressions through insertion, deletion, and replacement. For example,
they can replace relational, conditional, mathematical, or numerical
expressions, change transition sources or destinations, delete states or
transitions, remove state variables or transition conditions, and insert
new variables, mathematical operations, or transition conditions~\cite{arrieta2026flowrepair}.
These mutation operators allow FlowRepair to explore different types
of repairs while remaining specific to the structure and semantics of
Stateflow models.

The repair process continues until the defined time budget is reached.
During this process, every candidate patch is evaluated through
simulation. Candidate patches that pass the available test suite are
stored as plausible patches. However, a plausible patch is not
automatically guaranteed to be a valid patch, because it may overfit the
available tests and still be semantically incorrect. Therefore, FlowRepair
returns the plausible patches to the engineer for manual inspection and
validation. A valid patch is the one that not only passes the test suite
but also correctly fixes the underlying fault according to the intended
model behavior~\cite{arrieta2026flowrepair}.

\section{LLMs as Mutators in FlowRepair} 
\label{sec:LLMasmutators}

In this paper, we investigate more flexible label-level mutation generation. To this end, we replace some of the mutation operators from FlowRepair~\cite{arrieta2026flowrepair} with an LLM-based mutation operator. Table~\ref{table:MutOperators} explains the 15 mutation operators that the original FlowRepair~\cite{arrieta2026flowrepair} paper included. Given a buggy Stateflow model, these operators could be applied either to Transitions, States, or both. Our new tool substitutes some of these operators with a generic LLM that changes either the transition or the state.
%Some of the operators, however, need to be maintained
However, operators that change the structure of the Stateflow have been preserved. These include (i) Transition Destination Replacement, (ii) Transition Root Replacement, (iii) Initial Transition Change, (iv) State Deletion and (v) Transition Deletion. The rationale for this is that, in order to enable a more targeted repair, the only input we provide to the LLM is the specific content that a State or a Transition has. Subsequently, the LLM cannot make structural modifications to the Stateflow model. Therefore, we had to maintain these five operators to repair potential bugs coming from buggy structural decisions (e.g., wrong initial state).

\begin{table*}[!t]
\caption{Description and application for the 15 selected mutation operators. R, D and I refer to replacement, deletion and insertion, respectively.}
\label{table:MutOperators}
\centering
\resizebox{\textwidth}{!}{
\begin{tabular}{l|l|l|l}
\toprule
 
\multicolumn{1}{c|}{\textbf{Mutation Operator}} & 
\multicolumn{1}{c|}{\textbf{Description}} & 
\multicolumn{1}{c|}{\textbf{Application}} & 
\multicolumn{1}{c}{\textbf{Our new tool}} \\ \hline

\begin{tabular}[c]{@{}l@{}}Relational Operator \\ Replacement (R)\end{tabular} 
& \begin{tabular}[c]{@{}l@{}}Replaces a relational operator with another\\ (e.g., $>$ for $\geq$).\end{tabular} 
& Transitions 
& Substituted by generic LLM \\ \hline

\begin{tabular}[c]{@{}l@{}}Conditional Operator \\ Replacement (R)\end{tabular} 
& \begin{tabular}[c]{@{}l@{}}Replaces a conditional operator with another \\ (e.g., \texttt{\&\&} for \texttt{||}).\end{tabular} 
& Transitions 
& Substituted by generic LLM  \\ \hline

\begin{tabular}[c]{@{}l@{}}Mathematical Operator \\ Replacement (R)\end{tabular} 
& \begin{tabular}[c]{@{}l@{}}Replaces a mathematical operator with another \\ (e.g., $+$ for $-$).\end{tabular} 
& \begin{tabular}[c]{@{}l@{}}Transitions \\ and States\end{tabular} 
& Substituted by generic LLM  \\ \hline

\begin{tabular}[c]{@{}l@{}}Unit Change in After \\ Function (R)\end{tabular} 
& \begin{tabular}[c]{@{}l@{}}Changes the time unit in an after function \\ (e.g., \texttt{after(5, sec)} for \texttt{after(5, msec)}).\end{tabular} 
& Transitions 
& Substituted by generic LLM  \\ \hline

\begin{tabular}[c]{@{}l@{}}Numerical Replacement \\ Operator (R)\end{tabular} 
& Changes a numeric value for another (e.g., 1 for -1). 
& \begin{tabular}[c]{@{}l@{}}Transitions \\ and States\end{tabular} 
& Substituted by generic LLM \\ \hline

\begin{tabular}[c]{@{}l@{}}Transition Destination \\ Replacement (R)\end{tabular} 
& \begin{tabular}[c]{@{}l@{}}Changes the destination state of a transition to another\\ randomly chosen state.\end{tabular} 
& Transitions 
& Maintained \\ \hline

\begin{tabular}[c]{@{}l@{}}Transition Root \\ Replacement (R)\end{tabular} 
& \begin{tabular}[c]{@{}l@{}}Changes the origin state of a transition to another \\ randomly chosen state.\end{tabular} 
& Transitions 
& Maintained \\ \hline

\begin{tabular}[c]{@{}l@{}}Initial Transition \\ Change (R)\end{tabular} 
& Changes the initial transition. 
& Initial Transitions 
& Maintained \\ \hline

State Deletion (D) 
& \begin{tabular}[c]{@{}l@{}}Deletes a state from the Stateflow model; removes \\ affected transitions as well.\end{tabular} 
& States 
& Maintained \\ \hline

Transition Deletion (D) 
& Deletes a transition from the Stateflow model. 
& Transitions 
& Maintained \\ \hline

State Variable Deletion (D) 
& Deletes a variable from a state. 
& States 
& Substituted by generic LLM \\ \hline

\begin{tabular}[c]{@{}l@{}}Transition Condition \\ Deletion (D)\end{tabular} 
& \begin{tabular}[c]{@{}l@{}}Deletes a condition from a transition; applicable only \\ to transitions with multiple conditions.\end{tabular} 
& Transitions 
& Substituted by generic LLM  \\ \hline

\begin{tabular}[c]{@{}l@{}}Mathematical Operation \\ Insertion (I)\end{tabular} 
& \begin{tabular}[c]{@{}l@{}}Inserts a mathematical operation based on the model's \\ inputs and outputs into a targeted transition.\end{tabular} 
& \begin{tabular}[c]{@{}l@{}}Transitions \\ and States\end{tabular} 
& Substituted by generic LLM  \\ \hline

Variable Insertion (I) 
& Inserts a new variable in a state. 
& States 
& Substituted by generic LLM  \\ \hline

Condition Insertion (I) 
& Inserts a new condition in a transition. 
& Transitions 
& Substituted by generic LLM  \\ \bottomrule

\end{tabular}}
\end{table*}

We carefully craft two prompts, one for mutating states and the other one for mutating transitions. Fig.~\ref{fig:llm-prompt} shows the prompt template used to generate LLM-based mutations. The prompts get enough context to understand the Stateflow model, including input variables, output variables and internal variables. In addition, each prompt includes the name and current label of the selected component, along with a mutation description that specifies the repair instructions. The mutation description provides the LLM with representative few-shot examples of possible modifications, such as replacing boolean values ($\text{true} \leftrightarrow \text{false}$) or removing duplicated expressions. However, the LLM is not limited to these specific changes and can generate other suitable repairs.

\begin{figure}[t]
\centering
\setlength{\fboxsep}{5pt}
\fbox{%
\begin{minipage}{0.92\columnwidth}
\scriptsize
\raggedright

\textbf{Role.} You are a Stateflow repair assistant. Your task is to repair labels in a Stateflow model.

\medskip
\textbf{Inputs.}\\
Component type: \texttt{<STATE | TRANSITION>}\\
Component identifier: \texttt{<STATE\_NAME | TRANSITION\_ID>}\\
Current label: \texttt{<CURRENT\_LABEL>}\\
Available variables: \texttt{<VARIABLE\_LIST>}

\medskip
\textbf{Common constraints.}
Use only the provided variables, preserve valid MATLAB/Stateflow syntax, and return one candidate mutation in strict JSON format.

\medskip
\textbf{State-specific repair instructions.}
For state labels, preserve the \texttt{entry}, \texttt{during}, and \texttt{exit} structure when present. The mutation may replace numeric or Boolean values, modify assignment expressions, or insert/delete simple variable assignments.

\medskip
\textbf{Transition-specific repair instructions.}
For transition labels, the mutation may replace relational, logical, conditional, or mathematical operators; change numeric values or time delays such as \texttt{after(...)}; remove duplicated expressions; simplify or expand compound conditions.

\medskip
\textbf{Restrictions.}
Do not invent variables, change source or destination states, delete transitions, or add complex nested conditions.

\medskip
\textbf{Output format.}\\
Return strict JSON only, with no explanations or extra text:\\
\texttt{\{"patch": "<patched label>"\}}

\medskip
\textbf{Few-shot examples.}\\
\textit{State example:}\\
Input label: \texttt{entry: HOT = 1;}\\
Candidate patch: \texttt{\{"patch": "entry: HOT = 0;"\}}

\medskip
\textit{Transition example:}\\
Input label: \texttt{[TEMP > REF + 2 || TEMP < REF + 2]}\\
Candidate patch: \texttt{\{"patch": "[TEMP > REF + 2]"\}}

\medskip

\end{minipage}%
}
\caption{Simplified prompt template for LLM-based patch generation.}
\label{fig:llm-prompt}
\end{figure}

Furthermore, the prompts impose explicit constraints, requiring the LLM to use only the provided variables, thereby preventing the introduction of undeclared variables or unsupported constructs. The specific prompt we employ and the complete few-shot examples used in the experiments can be found in the replication package.

\section{Empirical Evaluation}
\label{sec:evaluation}
\subsection{Research Questions}

In our evaluation, we aim to answer the following research questions (RQs):

\vspace{6pt}

\begin{itemize}
    \setlength{\itemsep}{10pt}
    \setlength{\leftskip}{6pt}

    \item \textbf{\textit{RQ1 – Repairing ability:}}
    \textit{To what extent can the proposed LLM-based repair approach repair Stateflow faults?} This RQ examines the capability of the approach in producing plausible and valid patches to repair faults in Stateflow models.

    \item \textbf{\textit{RQ2 – Comparison with other algorithms:}}
    \textit{How does the proposed LLM-based repair approach perform in comparison to the original \textsc{FlowRepair} algorithm?} This RQ analyzes the impact of replacing rule-based mutation synthesis with LLM-guided generation, highlighting strengths, limitations, and performance differences relative to \textsc{FlowRepair}.

\end{itemize}

\subsection{Experimental Configuration}
\textbf{Overall setup: }The experimental foundation of this study follows the setup established in the prior work by Arrieta et al.~\cite{arrieta2026flowrepair}, which we adopt to ensure a controlled comparison under the same wall-clock budget between randomized and LLM-based mutation strategies. We use the same dataset and bug benchmarks, along with an identical preprocessing pipeline for fault localization, suspiciousness score calculation, and evaluation metrics defined in their study. Furthermore, all experiments are executed within the same infrastructure and computational environment described in the FlowRepair paper~\cite{arrieta2026flowrepair}. By reusing this setup rather than redefining it, we maintain methodological consistency and focus our contribution on analyzing how LLM‑driven repair behaves under the same controlled conditions.

\textbf{Algorithm Configuration: }The repair process was executed with a time budget of 1 hour and a maximum of 30 local repair attempts on partially improved patches, these values remain constant across all runs. Under this configuration, each run produced between 62 and 675 candidate patches with GPT-5.5, between 56 and 615 with GPT-5.4-mini, and between 75 and 713 with GPT-4.1-mini. The fault localization was guided using the Tarantula suspiciousness metric~\cite{wong2016faultlocalization}, one of the most widely used SBFL techniques. 

\textbf{Evaluation Metrics: }As in~\cite{arrieta2026flowrepair}, for each execution, two evaluation metrics were collected: the number of generated \textit{plausible patches}, defined as patches that pass the simulation-based test suite, and the number of generated \textit{valid patches}, i.e., plausible patches manually inspected by a software engineer and judged to implement the same corrective behavior as the developer patch. The Mean, minimum, and maximum summarize patch occurrences across five independent executions, where each generated patch is counted as one occurrence, including repeated identical patches. Therefore, these metrics quantify patch generation frequency rather than patch diversity.

% i.e., patches that are semantically equivalent to the fix provided by the developer.

\textbf{Execution runs: }Due to the stochastic nature of the repair process, the experiments were repeated five times for each faulty model. Taking into account 19 faulty models and a time budget of 1 hour, the total experimental effort amounted to 95 hours (5 $\times$ 19) per LLM variant. A higher number of repetitions was not feasible due to both the high computational cost and the manual effort required to validate the generated plausible patches, which involved manually validating a total of 1,724 plausible patch occurrences for semantic equivalence: 232 for GPT-5.5, 1,017 for GPT-5.4-mini, and 475 for GPT-4.1-mini.

%inspecting a total of 1724 generated plausible patch occurrences for semantic equivalence, including 232 plausible patches for GPT-5.5, 1017 for GPT-5.4-mini, and 475 for GPT-4.1-mini. %inspecting 601 plausible patches for semantic equivalence. 

\textbf{Employed LLM: }
We used GPT-5.5 as the initial model for the LLM-based repair because it offers strong reasoning ability for handling semantic changes in Stateflow labels. To assess whether the observed results are robust across different model choices, we additionally evaluated GPT-5.4-mini and GPT-4.1-mini. These smaller variants offer a better balance between performance and computational cost, allowing us to execute many repair attempts within the fixed time budget and repeat experiments across multiple models. The additional experiments help determine whether the observed repair behavior is specific to GPT-5.5 or remains consistent across different LLM variants. For patch generation, we used a fixed temperature of 0.8 for all LLM configurations to encourage diverse patch generation while maintaining a controlled comparison across models, consistent with other LLM-based APR studies~\cite{Fan2023Automated}.
Since the objective of this study is to evaluate LLM-generated mutations as a direct substitute for FlowRepair’s domain-specific operators, keeping the decoding configuration fixed avoids confounding the comparison with additional parameter tuning.

%For patch generation, we set the temperature to 0.8 to encourage diverse patch generation, helping the repair process explore different possible fixes, similar to other APR studies that use LLMs~\cite{Fan2023Automated}. We used a fixed temperature of for all LLM configurations to maintain a controlled comparison across models within the same repair budget and pipeline.
%We used GPT-4.1-mini for the LLM-based repair because it offers a good balance between performance and computational cost. Compared to larger models, it offers sufficient reasoning ability to handle semantic changes in Stateflow labels, allowing us to execute many repair attempts within the fixed time budget and repeat experiments across multiple models. 

\textbf{Dataset of real bugs:} The evaluation of the LLM-based repair approach was conducted by replicating the original FlowRepair paper, using the same dataset originally introduced in FlowRepair~\cite{arrieta2026flowrepair}, which consists of 19 faulty Stateflow models from four different case-studies. Table~\ref{tab:stateflow-characteristics} summarizes the structural characteristics of the Stateflow case-study models.
\begin{table}[t]
\centering
\caption{Structural characteristics of the Stateflow case studies}
\label{tab:stateflow-characteristics}
\scriptsize
\setlength{\tabcolsep}{3pt}
\begin{tabular}{lccccc}
\toprule
\textbf{Case Study} & 
\textbf{\# States} & 
\textbf{\# Trans.} & 
\textbf{\# Inputs} & 
\textbf{\# Outputs} & 
\textbf{\# Junc.} \\
\midrule
pacemaker & 15 & 25 & 10 & 14 & 0 \\
fridge    & 4  & 8  & 3  & 3  & 0 \\
door      & 6  & 10 & 6  & 1  & 0 \\
elevator  & 5  & 12 & 1  & 3  & 3 \\
\bottomrule
\end{tabular}
\end{table}

\subsection{Analysis of the Results}
%In this section we analyze the effectiveness of the proposed LLM-based repair approach in producing plausible and valid patches and address the research questions. %First, we examine the extent to which the LLM-based approach can produce plausible and valid repairs across a set of 19 faulty models. Second, the results of LLM based repair re compared with FlowRepair to analyze the differences in patch generation between the two approaches.
\subsubsection{RQ1- Repairing ability of the LLM-based approach}
The results of our evaluation are summarized in Table~\ref{tab:plausible-results} and Table~\ref{tab:valid_results}. Table~\ref{tab:plausible-results} reports the plausible patch results, while Table~\ref{tab:valid_results} presents the corresponding valid patch results across five independent runs. Across the five executions performed for each faulty model, the LLM-based repair approach showed limited repair ability across all evaluated LLM variants. GPT-5.5 generated plausible and valid patches for 4 out of the 19 models. GPT-5.4-mini generated plausible patches for 6 models and valid patches for 4 models, while GPT-4.1-mini generated plausible patches for 5 models and valid patches for 4 models. These results indicate that changing the LLM model affects the number of generated patches and the specific faults repaired, but the overall repairing ability remains limited under the evaluated conditions.

\begin{table*}[t]
\centering
\caption{Plausible patch results across 5 runs comparing FlowRepair and the LLM-based repair variants.}
\label{tab:plausible-results}
\scriptsize
\setlength{\tabcolsep}{4pt}
\renewcommand{\arraystretch}{1.05}

\begin{tabular}{l
  *{3}{>{\centering\arraybackslash}p{0.55cm}} |
  *{3}{>{\centering\arraybackslash}p{0.55cm}} |
  *{3}{>{\centering\arraybackslash}p{0.55cm}} |
  *{3}{>{\centering\arraybackslash}p{0.55cm}}
}
\toprule
\textbf{Model}
& \multicolumn{3}{c|}{\textbf{FlowRepair}}
& \multicolumn{3}{c|}{\makecell{\textbf{LLM Repair}\\\textbf{(GPT-5.5)}}}
& \multicolumn{3}{c|}{\makecell{\textbf{LLM Repair}\\\textbf{(GPT-5.4-mini)}}}
& \multicolumn{3}{c}{\makecell{\textbf{LLM Repair}\\\textbf{(GPT-4.1-mini)}}} \\
\cmidrule(lr){2-4}
\cmidrule(lr){5-7}
\cmidrule(lr){8-10}
\cmidrule(lr){11-13}
& \textbf{Mean} & \textbf{Min} & \textbf{Max}
& \textbf{Mean} & \textbf{Min} & \textbf{Max}
& \textbf{Mean} & \textbf{Min} & \textbf{Max}
& \textbf{Mean} & \textbf{Min} & \textbf{Max} \\
\midrule

\textbf{pacemaker\_1}  & \textbf{4.2}  & 0  & \textbf{11} & 0.6  & 0  & 2  & 0    & 0   & 0   & 0    & 0  & 0  \\
\textbf{pacemaker\_2}  & 0    & 0  & 0  & \textbf{14}   & \textbf{9}  & \textbf{21} & 0    & 0   & 0   & 0    & 0  & 0  \\
\textbf{fridge\_1}     & 8.8  & 3  & 14 & 0    & 0  & 0  & \textbf{35.2} & \textbf{7}   & \textbf{50}  & 0.6  & 0  & 2  \\
\textbf{fridge\_2}     & 4.2  & 0  & 8  & 0    & 0  & 0  & 0    & 0   & 0   & 0    & 0  & 0  \\
\textbf{fridge\_2a}    & 18.4 & 15 & 24 & 20.2 & 13 & 31 & \textbf{142.8}& \textbf{130} & \textbf{154} & 63.2 & 51 & 75 \\
\textbf{fridge\_2b}    & \textbf{18.8} & \textbf{15} & \textbf{25} & 0    & 0  & 0  & 0    & 0   & 0   & 0    & 0  & 0  \\
\textbf{fridge\_3}     & \textbf{1.0}  & \textbf{1}  & \textbf{1}  & 0    & 0  & 0  & 0    & 0   & 0   & 0    & 0  & 0  \\
\textbf{door\_1}       & \textbf{13.4} & \textbf{8}  & \textbf{21} & 0    & 0  & 0  & 0    & 0   & 0   & 1.6  & 0  & 3  \\
\textbf{door\_2}       & \textbf{9.8}  & \textbf{6}  & \textbf{15} & 0    & 0  & 0  & 0    & 0   & 0   & 0    & 0  & 0  \\
\textbf{elevator\_1}   & \textbf{0.8}  & 0  & \textbf{2}  & 0    & 0  & 0  & 0    & 0   & 0   & 0    & 0  & 0  \\
\textbf{elevator\_2}   & \textbf{1.8}  & 0  & \textbf{5}  & 0    & 0  & 0  & 0    & 0   & 0   & 0    & 0  & 0  \\
\textbf{elevator\_3}   & \textbf{6.8}  & \textbf{4}  & \textbf{13} & 0    & 0  & 0  & 0    & 0   & 0   & 0    & 0  & 0  \\
\textbf{elevator\_4}   & \textbf{14.8} & \textbf{7}  & \textbf{20} & 0    & 0  & 0  & 1.8  & 1   & 4   & 0    & 0  & 0  \\
\textbf{elevator\_5}   & \textbf{19.2} & \textbf{14} & \textbf{24} & 0    & 0  & 0  & 18.6 & 14  & 25  & 0    & 0  & 0  \\
\textbf{elevator\_6}   & \textbf{8.4}  & \textbf{6}  & \textbf{13} & 0    & 0  & 0  & 0    & 0   & 0   & 0    & 0  & 0  \\
\textbf{elevator\_7}   & 1.0  & 0  & 2  & 0    & 0  & 0  & 2.4  & 0   & 5   & \textbf{17.8} & \textbf{12} & \textbf{21} \\
\textbf{elevator\_8}   & 1.4  & 0  & 3  & 11.6 & 2  & 26 & 2.6  & 1   & 4   & \textbf{11.8} & \textbf{10} & \textbf{14} \\
\textbf{elevator\_9}   & \textbf{3.4}  & \textbf{2}  & \textbf{6}  & 0    & 0  & 0  & 0    & 0   & 0   & 0    & 0  & 0  \\
\textbf{elevator\_10}  & \textbf{0.6}  & 0  & \textbf{2}  & 0    & 0  & 0  & 0    & 0   & 0   & 0    & 0  & 0  \\

\bottomrule
\end{tabular}
\end{table*}

\begin{table*}[t]
\centering
\caption{Valid patch results across 5 runs comparing FlowRepair and the LLM-based repair variants.}
\label{tab:valid_results}
\scriptsize
\setlength{\tabcolsep}{4pt}
\renewcommand{\arraystretch}{1.05}

\begin{tabular}{l
  *{3}{>{\centering\arraybackslash}p{0.55cm}} |
  *{3}{>{\centering\arraybackslash}p{0.55cm}} |
  *{3}{>{\centering\arraybackslash}p{0.55cm}} |
  *{3}{>{\centering\arraybackslash}p{0.55cm}}
}
\toprule
\textbf{Model}
& \multicolumn{3}{c|}{\textbf{FlowRepair}}
& \multicolumn{3}{c|}{\makecell{\textbf{LLM Repair}\\\textbf{(GPT-5.5)}}}
& \multicolumn{3}{c|}{\makecell{\textbf{LLM Repair}\\\textbf{(GPT-5.4-mini)}}}
& \multicolumn{3}{c}{\makecell{\textbf{LLM Repair}\\\textbf{(GPT-4.1-mini)}}} \\
\cmidrule(lr){2-4}
\cmidrule(lr){5-7}
\cmidrule(lr){8-10}
\cmidrule(lr){11-13}
& \textbf{Mean} & \textbf{Min} & \textbf{Max}
& \textbf{Mean} & \textbf{Min} & \textbf{Max}
& \textbf{Mean} & \textbf{Min} & \textbf{Max}
& \textbf{Mean} & \textbf{Min} & \textbf{Max} \\
\midrule

\textbf{pacemaker\_1}  & 0.4  & 0  & 1  & \textbf{0.6}  & 0  & \textbf{2}  & 0    & 0   & 0   & 0    & 0  & 0  \\
\textbf{pacemaker\_2}  & 0    & 0  & 0  & \textbf{14}   & \textbf{9}  & \textbf{21} & 0    & 0   & 0   & 0    & 0  & 0  \\
\textbf{fridge\_1}     & 8.8  & 3  & 14 & 0    & 0  & 0  & \textbf{35.2} & \textbf{7}   & \textbf{50}  & 0.6  & 0  & 2  \\
\textbf{fridge\_2}     & \textbf{0.4}  & 0  & \textbf{2}  & 0    & 0  & 0  & 0    & 0   & 0   & 0    & 0  & 0  \\
\textbf{fridge\_2a}    & 13.4 & 9  & 17 & 20.2 & 13 & 31 & \textbf{142.8} & \textbf{130} & \textbf{154} & 63.2 & 51 & 75 \\
\textbf{fridge\_2b}    & \textbf{7.4}  & \textbf{4}  & \textbf{9}  & 0    & 0  & 0  & 0    & 0   & 0   & 0    & 0  & 0  \\
\textbf{fridge\_3}     & 0    & 0  & 0  & 0    & 0  & 0  & 0    & 0   & 0   & 0    & 0  & 0  \\
\textbf{door\_1}       & 0    & 0  & 0  & 0    & 0  & 0  & 0    & 0   & 0   & 0    & 0  & 0  \\
\textbf{door\_2}       & \textbf{2}    & \textbf{0}  & \textbf{6}  & 0    & 0  & 0  & 0    & 0   & 0   & 0    & 0  & 0  \\
\textbf{elevator\_1}   & \textbf{0.8}  & 0  & \textbf{2}  & 0    & 0  & 0  & 0    & 0   & 0   & 0    & 0  & 0  \\
\textbf{elevator\_2}   & \textbf{1.8}  & 0  & \textbf{5}  & 0    & 0  & 0  & 0    & 0   & 0   & 0    & 0  & 0  \\
\textbf{elevator\_3}   & \textbf{6.8}  & \textbf{4}  & \textbf{13} & 0    & 0  & 0  & 0    & 0   & 0   & 0    & 0  & 0  \\
\textbf{elevator\_4}   & \textbf{14.8} & \textbf{7}  & \textbf{20} & 0    & 0  & 0  & 0    & 0   & 0   & 0    & 0  & 0  \\
\textbf{elevator\_5}   & \textbf{19.2} & \textbf{14} & \textbf{24} & 0    & 0  & 0  & 0    & 0   & 0   & 0    & 0  & 0  \\
\textbf{elevator\_6}   & \textbf{8.4}  & \textbf{6}  & \textbf{13} & 0    & 0  & 0  & 0    & 0   & 0   & 0    & 0  & 0  \\
\textbf{elevator\_7}   & 1.0  & 0  & 2  & 0    & 0  & 0  & 2.4  & 0   & 5   & \textbf{17.8} & \textbf{12} & \textbf{21} \\
\textbf{elevator\_8}   & 1.0  & 0  & 2  & 11.6 & 2  & 26 & 2.6  & 1   & 4   & \textbf{11.8} & \textbf{10} & \textbf{14} \\
\textbf{elevator\_9}   & \textbf{3.4}  & \textbf{2}  & \textbf{6}  & 0    & 0  & 0  & 0    & 0   & 0   & 0    & 0  & 0  \\
\textbf{elevator\_10}  & \textbf{0.6}  & 0  & \textbf{2}  & 0    & 0  & 0  & 0    & 0   & 0   & 0    & 0  & 0  \\

\bottomrule
\end{tabular}
\end{table*}

\begin{figure*}[t]
\centering

\begin{subfigure}[b]{0.38\textwidth}
    \centering
    \includegraphics[width=\linewidth]{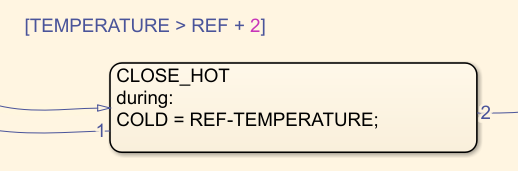}
    \caption{Fault in fridge\_1}
    \label{fig:fridge1_fault}
\end{subfigure}
\hspace{0.04\textwidth}
\begin{subfigure}[b]{0.38\textwidth}
    \centering
    \includegraphics[width=\linewidth]{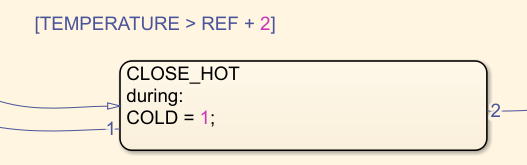}
    \caption{Patch for fridge\_1}
    \label{fig:fridge1_patch}
\end{subfigure}

\vspace{0.3em}

\begin{subfigure}[b]{0.38\textwidth}
    \centering
    \includegraphics[width=\linewidth]{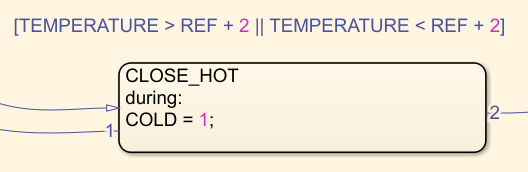}
    \caption{Fault in fridge\_2a}
    \label{fig:fridge2a_fault}
\end{subfigure}
\hspace{0.04\textwidth}
\begin{subfigure}[b]{0.38\textwidth}
    \centering
    \includegraphics[width=\linewidth]{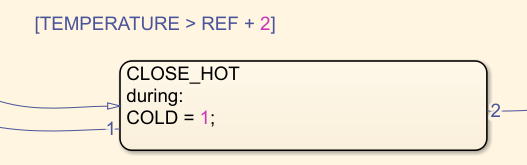}
    \caption{Patch for fridge\_2a}
    \label{fig:fridge2a_patch}
\end{subfigure}

\vspace{-0.5em}
\caption{Faults and corresponding patches for fridge models.}
\label{fig:fridge2a}
\vspace{-1em}
\end{figure*}

A closer inspection of the successful cases shows that the LLM-based approach is most effective when the fault can be repaired through localized semantic transformations in Stateflow labels. For example, in the fridge\_2a model, the fault involves a redundant and logically inconsistent condition in a transition guard. Repairing this issue primarily requires simplifying the expression by removing the unnecessary clause. This model was repaired by all evaluated LLM variants, although with different numbers of patches. As the fridge\_2a repair pattern was present in the few-shot prompt, this result represents prompt-conditioned generation: GPT-5.5 produced a mean of 20.2 plausible and valid patches, GPT-5.4-mini produced 142.8, and GPT-4.1-mini produced 63.2. Similarly, elevator\_8 was repaired by all three variants, while elevator\_7 was repaired by GPT-5.4-mini and GPT-4.1-mini. These results suggest that LLM-generated mutations can be useful when the correct repair lies close to the original expression and can be obtained through a small local modification. Figure~\ref{fig:fridge2a} illustrates both faults and the corresponding patches suggested by our approach.

The results also show that successful repair cases vary across model choices. For instance, GPT-5.5 repaired both Pacemaker models, which were not repaired by GPT-5.4-mini or GPT-4.1-mini. Conversely, GPT-5.4-mini repaired fridge\_1 with a mean of 35.2 plausible and valid patches, whereas GPT-5.5 did not repair this model and GPT-4.1-mini produced only a small number of patches. This variation suggests that some repair outcomes are model-dependent. However, the broader trend remains consistent: across the evaluated LLM variants, successful repairs are concentrated in a small subset of faults.

In contrast, the approach performs poorly for faults that require precise value synthesis, missing expression insertion, or structural reasoning. For example, door\_2 requires changing a specific assigned value from 1 to -1, while elevator\_1 and elevator\_2 require inserting missing expressions. None of the evaluated LLM variants produced valid patches for these models. Similarly, the approach struggles with structural faults such as door\_1, where the correct repair requires modifying the source of a transition. GPT-4.1-mini generated plausible patches for this model, but none were valid, suggesting that the generated changes passed the available tests without capturing the intended structural correction. A similar pattern appears in elevator\_4 and elevator\_5, where GPT-5.4-mini generated plausible patches but no valid patches, indicating overfitting or semantically incorrect repairs.

Overall, these results indicate that the effectiveness of LLM-based mutation depends strongly on the nature of the fault and the selected LLM variant. The approach can generate correct repairs for some localized semantic changes, but it remains unreliable for faults requiring exact value synthesis, missing expressions, or structural modifications. The additional experiments with GPT-5.4-mini and  GPT-4.1-mini strengthen this observation by showing that, although the exact successful cases differ across models, the limited repair coverage remains consistent across the evaluated LLM choices.

After analyzing the results, we can summarize the first RQ as follows:\newline

\noindent
\fbox{%
\parbox{0.98\linewidth}{%
\textbf{RQ1:} Using LLMs as repair mutators produced plausible patches for 4 out of 19 models with GPT-5.5, 6 out of 19 with GPT-5.4-mini, and 5 out of 19 with GPT-4.1-mini. Valid patches were produced for 4 out of 19 models for each evaluated LLM variant. These results indicate limited repairing ability under the evaluated conditions, with successful repairs mainly concentrated in localized semantic faults.}}
\vspace{0.8em}

\subsubsection{RQ2 – Comparison with FlowRepair~\cite{arrieta2026flowrepair}} 
To assess the effectiveness of the proposed approach, we compare the LLM-based repair variants against the original FlowRepair technique. The results, reported in Tables~\ref{tab:plausible-results} and~\ref{tab:valid_results}, show a clear performance gap between FlowRepair and all evaluated LLM-based variants. Across the evaluated models and five independent runs, FlowRepair generated plausible patches for 18 out of 19 faulty models and valid patches for 16 models. In comparison, all evaluated LLM-based variants repaired only a limited subset of the benchmark's faults, with valid patches produced for 4 models in each variant and plausible patch coverage ranging from 4 to 6 models. These results indicate that, under the evaluated repair configuration, FlowRepair remains more effective and consistent than the tested LLM-based variants.

The stronger performance of FlowRepair may be explained by its use of carefully designed Stateflow-specific mutation operators combined with a guided search strategy. These operators generate syntactically valid and semantically constrained modifications, which preserve a more focused search space and help the algorithm converge toward valid repairs. In contrast, the LLM-based variants replace several of these operators with generative mutations. Although this increases flexibility, it also expands the search space and may reduce the proportion of useful candidates. As a result, many generated patches either do not pass simulation-based validation or pass the available tests without being semantically valid.

Despite their overall weaker performance, the LLM-based variants show competitive or superior results in a small subset of models. GPT-5.5 successfully repaired pacemaker\_1, pacemaker\_2, fridge\_2a, and elevator\_8, showing that a stronger model can repair some faults that were not repaired by the smaller variants. For pacemaker\_1, GPT-5.5 produced fewer plausible patches than FlowRepair on average, but a slightly higher number of valid patches. For pacemaker\_2, GPT-5.5 produced a mean of 14 plausible and valid patches, whereas FlowRepair, GPT-5.4-mini, and GPT-4.1-mini produced no plausible or valid patches. GPT-5.4-mini achieved the strongest results for fridge\_1 and fridge\_2a, producing higher numbers of plausible and valid patches than FlowRepair in these cases. GPT-4.1-mini also performed well on fridge\_2a, elevator\_7, and elevator\_8. These successful cases mainly involve localized semantic transformations, such as simplifying transition guards, correcting arithmetic operators, or replacing state-label expressions. In such cases, LLM-generated mutations can explore useful alternatives beyond predefined mutation choices.

However, these advantages are not consistent across the benchmark. In the majority of models, especially those requiring structural modifications, missing expression insertion, or precise value synthesis, FlowRepair remains more reliable. For example, the LLM-based variants fail to produce valid patches for door\_1, where the correct repair requires changing the transition source, and for elevator\_1 and elevator\_2, where the fix requires inserting missing expressions. GPT-5.4-mini also generates plausible but not valid patches for elevator\_4 and elevator\_5, suggesting that some generated repairs satisfy the available tests without matching the intended fix. FlowRepair remains more robust on several fault types that the evaluated LLM variants did not repair. For example, FlowRepair generated plausible and valid patches for elevator\_3, elevator\_6, elevator\_9, and elevator\_10, with mean valid patch counts of 6.8, 8.4, 3.4, and 0.6, respectively, while all LLM-based variants produced zero plausible patches for these models. A similar pattern appears for fridge\_2b, where FlowRepair produced 18.8 plausible patches and 7.4 valid patches on average, but none of the LLM variants generated a plausible patch. For fridge\_2, FlowRepair produced 4.2 plausible patches and 0.4 valid patches on average, whereas the LLM variants again produced none. fridge\_3 is a weaker case for FlowRepair, as it produced plausible but not valid patches; however, the LLM variants still failed to generate any plausible patch. These cases suggest that the performance gap is not limited to one specific fault pattern, but appears across faults requiring coordinated edits, Stateflow-specific syntax, temporal semantics, and reliable operator-level mutation.%FlowRepair remains more robust on several fault types that the evaluated LLM variants did not repair. For example, Fridge\_2 requires coordinating two transition-label corrections, while Fridge\_2b and Elevator\_9 require localized operator replacements that were nevertheless not reached by the LLM-based search. Other missed cases involve more Stateflow-specific semantics: Fridge\_3 requires replacing a temporal condition with a temperature-based guard, Elevator\_6 requires removing an incorrectly added delay while preserving the intended transition condition, and Elevator\_10 requires correcting the syntax of an \texttt{after} expression. In addition, Elevator\_3 requires removing added exit actions from two states, which involves a coordinated state-action repair. These cases suggest that the performance gap is not limited to one specific fault pattern, but appears across faults requiring coordinated edits, Stateflow-specific syntax, temporal semantics, and reliable operator-level mutation.

Overall, the comparison highlights a trade-off between structured and generative mutation strategies in this Stateflow/CPS repair setting. The evaluated LLM-based variants can be beneficial for some localized semantic faults, but they do not consistently improve upon the domain-specific mutation operators used by FlowRepair. The additional results across GPT-5.5, GPT-5.4-mini, and GPT-4.1-mini also suggest that the observed performance gap is not limited to a single model choice, although the specific repaired faults vary across LLM variants.

We can therefore answer the second RQ as follows:\newline

\noindent
\fbox{%
\parbox{0.98\linewidth}{%
\textbf{RQ2:} Across the evaluated LLM variants, the LLM-based mutator variants are outperformed by FlowRepair in both repairing ability and consistency. While GPT-5.5, GPT-5.4-mini, and GPT-4.1-mini show advantages in a small number of cases involving localized semantic fixes, FlowRepair remains more reliable overall due to its structured Stateflow-specific mutation operators and guided search process.}}

\section{Discussion}
\label{sec:disc}

The results of our empirical evaluation show that replacing several hand-crafted mutation operators in FlowRepair with LLM-generated mutations did not improve repair performance for the evaluated Stateflow models. Across the 19 faulty models, 
FlowRepair achieved broader repair coverage, while the LLM-based variants repaired only a small subset, with successful cases concentrated in a few localized semantic faults.

%the original FlowRepair approach achieved substantially broader repair coverage than the evaluated LLM-based variants. While FlowRepair produced plausible and valid patches for most models, the LLM-based variants repaired only a small subset of the benchmark, with successful cases concentrated in a few localized semantic faults. The overall pattern indicates that, in this Stateflow/CPS repair setting, direct generative replacement of domain-specific mutation operators provides less reliable repair guidance than the structured operators used in FlowRepair.

\begin{figure*}[t]
\centering

\begin{subfigure}[b]{0.48\textwidth}
\centering
\includegraphics[width=\textwidth]{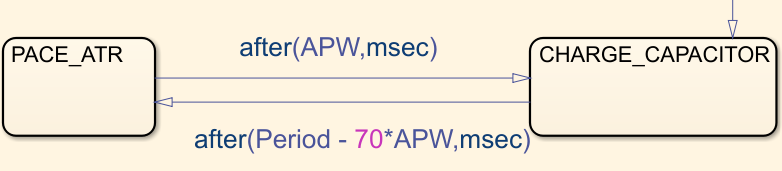}
\caption{Fault in pacemaker\_1}
\label{fig:pacemaker_1_fault}
\end{subfigure}
\hfill
\begin{subfigure}[b]{0.48\textwidth}
\centering
\includegraphics[width=\textwidth]{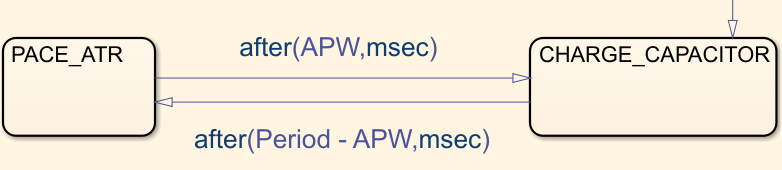}
\caption{Patch for pacemaker\_1}
\label{fig:pacemaker_1_patch}
\end{subfigure}

\vspace{0.6em}

\hspace{2em}
\begin{subfigure}[b]{0.36\textwidth}
\centering
\includegraphics[width=\textwidth]{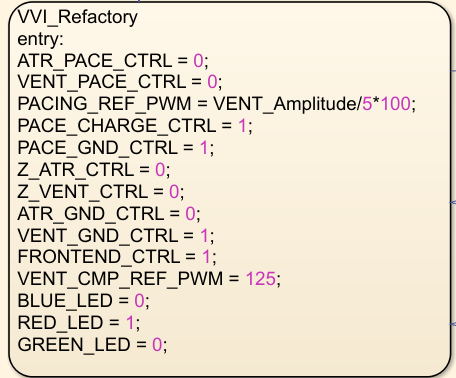}
\caption{Fault in pacemaker\_2}
\label{fig:pacemaker_2_fault}
\end{subfigure}
\hfill
\begin{subfigure}[b]{0.36\textwidth}
\centering
\includegraphics[width=\textwidth]{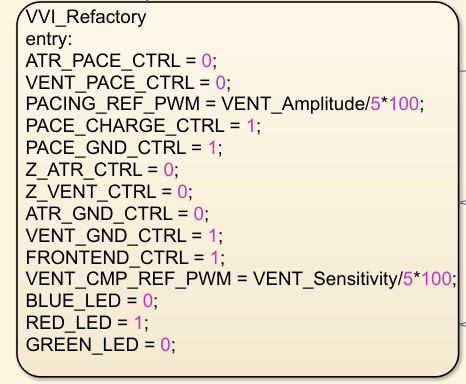}
\caption{Patch for pacemaker\_2}
\label{fig:pacemaker_2_patch}
\end{subfigure}
\hspace{2em}

\caption{Faults and corresponding patches for pacemaker models}
\label{fig:pacemaker_patches}
\end{figure*}

The exact repaired models differ across the three LLM variants, showing that model choice influences repair outcomes. For example, GPT-5.5 repaired pacemaker\_1, pacemaker\_2, fridge\_2a, and elevator\_8, whereas GPT-5.4-mini produced valid patches for fridge\_1, fridge\_2a, elevator\_7, and elevator\_8, while also generating plausible but invalid patches for elevator\_4 and elevator\_5. GPT-4.1-mini repaired fridge\_1, fridge\_2a, elevator\_7, and elevator\_8, and generated plausible but invalid patches for door\_1. GPT-5.5 was the only evaluated LLM variant that repaired pacemaker\_1 and pacemaker\_2. As shown in Figure~\ref{fig:pacemaker_patches}, in pacemaker\_1, the faulty transition label contains an additional multiplier in the timing expression, changing the intended delay from \texttt{after(Period - APW,msec)} to \texttt{after(Period - 70*APW,msec)}. In pacemaker\_2, the faulty state action assigns a fixed value, \texttt{VENT\_CMP\_REF\_PWM = 125;}, instead of the intended expression \texttt{VENT\_CMP\_REF\_PWM = VENT\_Sensitivity/5*100;}.
%In Pacemaker\_1, the faulty transition label contains an additional multiplier in the timing expression, changing the intended delay from \texttt{after(Period - APW,msec)} to \texttt{after(Period - 70APW,msec)}. In Pacemaker\_2, the faulty state action assigns a fixed value, \texttt{VENT_CMP_REF_PWM = 125;}, instead of the intended expression \texttt{VENT_CMP_REF_PWM = VENT_Sensitivity/5*100;}. 
These examples require recovering compact arithmetic expressions from the surrounding Stateflow context. The fact that GPT-5.5 repairs these cases while the smaller variants do not suggests that stronger reasoning and contextual rewriting ability can help in some narrow expression-repair scenarios. However, this advantage is not consistent across the benchmark, as GPT-5.5 still fails on several models repaired by FlowRepair. Overall, all three LLM variants repair only a limited subset of the benchmark and remain less consistent than FlowRepair.

%However, this advantage is not uniform across the benchmark, since GPT-5.5 still fails on several other models repaired by FlowRepair. Thus, stronger or newer models do not uniformly dominate smaller variants across all faults. Nevertheless, the aggregate trend is stable across model choices: all three LLM variants repair only a limited subset of the benchmark and remain less consistent than FlowRepair.

The LLM-based repair nevertheless shows strengths in specific cases, outperforming FlowRepair on fridge\_2a and achieving strong results on elevator\_7 and elevator\_8. These cases involve localized expression-level transformations, such as simplifying transition guards or modifying arithmetic expressions. However, the repair pattern for fridge\_2a was present in the few-shot prompt; thus, its successful repair represents prompt-conditioned generation rather than independent inference of the repair pattern. The elevator faults required small arithmetic edits, such as replacing $-$ with $+$ or simplifying $-1 + 1$ to $-1$. These repairs are close to the original expression and can benefit from the contextual rewriting ability of LLMs.

%The LLM-based repair nevertheless shows notable strength in a limited number of cases. In fridge\_2a, all three LLM variants outperform FlowRepair in terms of the number of plausible and valid patches. GPT-5.5 produced a mean of 20.2 plausible and valid patches, GPT-5.4-mini produced 142.8, and GPT-4.1-mini produced 63.2, compared to FlowRepair's mean of 18.4 plausible patches and 13.4 valid patches.Similarly, GPT-4.1-mini produced an average of 17.8 plausible and valid patches for elevator\_7, whereas FlowRepair produced only 1 plausible and valid patch on average. In elevator\_8, all LLM variants produced valid patches, with GPT-5.5 producing 11.6 valid patches on average, GPT-5.4-mini producing 2.6, and GPT-4.1-mini producing 11.8, compared to FlowRepair's average of 1 valid patch. These successful cases mainly involve localized expression-level transformations, such as simplifying a transition guard, correcting an arithmetic operator, or removing a redundant additive term. For example, fridge\_2a requires simplifying a compound condition from $[\text{TEMPERATURE} > \text{REF} + 2 || \text{TEMPERATURE} < \text{REF} + 2]$ to $[\text{TEMPERATURE} > \text{REF} + 2]$, while elevator\_7 and elevator\_8 require small arithmetic edits such as replacing $-$ with $+$ or simplifying $-1 + 1$ to $-1$. These repairs are close to the original expression and can benefit from the contextual rewriting ability of LLMs.

At the same time, not all localized or seemingly simple faults were repaired successfully. For example, elevator\_9 also requires an arithmetic operator correction, replacing $*$ with $+$ in a transition label, yet none of the evaluated LLM variants produced a plausible patch for this model. Similarly, fridge\_2b requires a local comparison-operator change from \texttt{==} to \texttt{>=}, but this repair was not reached by any of the LLM variants. Similarly, no plausible patches were generated for fridge\_2. This model combines two transition-label faults: one requiring simplification of a compound guard and another requiring correction of a comparison operator. Although the LLM variants repaired fridge\_2a when one of these faults was isolated, they did not repair the combined fridge\_2 model. This suggests that the evaluated LLM-based repair struggles when multiple related edits must be coordinated within the same repair run. These cases show that even small operator-level edits are not always generated reliably within the search process. Other missed cases require changes that go beyond direct expression rewriting. fridge\_3 is more semantically demanding because the incorrect label \texttt{[after(1,sec)]} must be replaced with a temperature-based guard, \texttt{[TEMPERATURE >= REF]}, requiring a shift from a temporal condition to a variable-based condition. elevator\_3 requires removing an added exit action from two states, which is a coordinated state-action repair rather than a single label rewrite. elevator\_6 requires removing an incorrectly added delay from a transition label while preserving the intended condition, which depends on Stateflow temporal semantics. Lastly, elevator\_10 requires removing brackets around an \texttt{after} expression, a small syntactic change with Stateflow-specific semantic implications. Together, these results suggest that repair success is not determined only by the syntactic size of the required edit. The surrounding Stateflow context, variable usage, generated candidate sequence, and interaction with the search process all influence whether useful candidates are produced within the available time budget.

%A first key insight concerns the shape of the search space. FlowRepair relies on deliberately constrained mutation operators that produce small, syntactically valid, and semantically local modifications. This design keeps the search space focused and increases the density of useful candidates. In contrast, LLM-based mutation introduces more flexible but less constrained edits. Although this may increase the variety of possible repairs, it also reduces the proportion of candidates that are useful for simulation-based validation. This effect is visible across the evaluated variants: each run generated many candidate patches, ranging from 62 to 675 with GPT-5.5, 56 to 615 with GPT-5.4-mini, and 75 to 713 with GPT-4.1-mini, depending on the model. Yet, only a small fraction of these candidates led to plausible or valid repairs.

A first key insight concerns the shape of the search space. FlowRepair relies on deliberately constrained mutation operators that produce small, syntactically valid, and semantically local modifications, keeping the search space focused. In contrast, LLM-based mutation introduces more flexible but less constrained edits. Although this may increase the variety of possible repairs, it may reduce the proportion of useful candidates for simulation-based validation. Each run generated many candidates, yet only a small fraction led to plausible or valid repairs. In CPS repair, where each candidate must be validated through simulation, such less constrained exploration may quickly consume the available repair budget. %noisy exploration can quickly consume the available repair budget.

A second factor is the precision demanded by many faults in the benchmark. Several models require exact value synthesis, missing expression insertion, or tightly constrained symbolic edits. For instance, door\_2 requires changing a specific assigned value from $1$ to $-1$, while elevator\_1 and elevator\_2 require inserting missing expressions. These faults were not repaired by any of the evaluated LLM variants. Similarly, fridge\_1 requires replacing the state action \texttt{COLD = REF-TEMPERATURE;} with the specific constant assignment \texttt{COLD = 1;}. GPT-5.4-mini performed well on this model, but GPT-5.5 failed to repair it and GPT-4.1-mini produced only a small number of valid patches. This indicates that even when the intended fix is conceptually simple, the set of valid alternatives can be narrow, making repair difficult when generation is not explicitly constrained by domain-specific operators.

Third, several failures arise from plausible but semantically invalid patches. This was observed in door\_1 for GPT-4.1-mini, and in elevator\_4 and elevator\_5 for GPT-5.4-mini, where plausible patches were generated but none were valid. These cases show that passing the available simulation tests does not always imply semantic correctness. In door\_1, the fault involves an incorrect transition source, which is a structural property of the Stateflow model. In elevator\_4 and elevator\_5, the fault involves an added exit action. These faults require changes that are tied to Stateflow execution semantics and model structure, rather than only local textual rewriting.

Structural characteristics of Stateflow models further compound these difficulties, as repairs may require graph-level changes, such as modifying transition sources or targets or adding/removing state actions while preserving execution order. The current LLM-based integration operates mainly on textual representations of states and transitions, providing limited access to topology and execution semantics. Although a subset of FlowRepair's structural operators was retained, the hybrid configuration did not match FlowRepair's reliability on structure-dependent faults, highlighting the importance of explicitly representing Stateflow structure in generative CPS repair.

%Structural characteristics of Stateflow models further compound these difficulties. Correct repairs may involve graph-level changes, such as modifying the source or target of a transition, or adding/removing state actions in a way that preserves execution order. The current LLM-based integration operates mainly on textual representations of states and transitions, giving the model limited access to topology and execution semantics. Although we retained a subset of structural operators from the original FlowRepair to preserve the possibility of structural repair, the results suggest that this hybrid design was not sufficient to match FlowRepair's reliability on structure-dependent faults. This points to the importance of representing Stateflow structure explicitly when using generative models for CPS repair.

The absence of behavioural feedback is another important design factor. FlowRepair's search is guided by repair objectives such as failure duration, severity, and activation time, which provide feedback after simulation. In the evaluated LLM-based variants, the LLM generates mutations from static information such as labels, actions, and variable lists, without receiving iterative feedback about why a patch failed or whether it partially improved the system behaviour. As a result, the generation process cannot directly refine its suggestions based on dynamic CPS behaviour. This could contribute to repeated generation of irrelevant, near-miss, or overfitting patches.

Taken together, these findings suggest that the effectiveness of search-based repair depends not only on the optimization algorithm, but also on the structure of the search space induced by the mutation operators. In this study, replacing several domain-specific operators with generative mutations reduced the effectiveness of the search process. At the same time, the successful cases show that LLM-generated mutations can be useful for specific localized semantic repairs. The key challenge is therefore to identify how generative models can complement domain-specific mutation mechanisms rather than replace them wholesale.

\textbf{Implications for the community: } Negative results like these are essential for calibrating our collective hypothesis space and research efforts. They caution against the uncritical transfer of LLM successes from general-purpose code repair to specialized, simulation-heavy domains. In CPS contexts, where each fitness evaluation is computationally expensive and test oracles are sparse, unconstrained generative mutations can actually reduce repair effectiveness. The study therefore reinforces the continuing value of carefully engineered, domain-aware operators while highlighting the need for more thoughtful integration strategies.

\textbf{Future directions:} A promising path forward lies in hybrid architectures that combine the complementary strengths of both paradigms. Possible designs include:

\begin{itemize}
\item Integrating simulation feedback into the prompting loop, e.g., ``Previous patch reduced failure duration by 40\%; suggest a refinement''.
\item Employing LLMs as repair advisors that rank, filter, or instantiate mutations generated by traditional operators rather than generating patches directly.
\item Selectively replacing only those mutation operators where LLMs show clear advantages, instead of replacing several operators at once.
\item Evaluating sensitivity to prompting strategy, temperature, and model choice on representative subsets of faults.
\item Fine-tuning or continued pre-training on Stateflow-specific corpora augmented with fault--fix pairs and execution traces~\cite{Guochang2024Automated}.
\end{itemize}

In conclusion, this replication and negative-results study shows that replacing hand-crafted mutation operators with LLM generation did not improve repair performance for CPS controllers modeled in Stateflow under the evaluated configuration. These findings suggest that combining the structured guidance of traditional search-based APR with the generative flexibility of LLMs may be a more promising direction.

\section{Threats to Validity}
\label{sec:threats}
\textbf{External Validity}: One of the threats to our evaluation relates to the generalization of the results. We employed 19 faulty models from four different case studies to mitigate this threat. Although the dataset size is not large compared to other domains, it is one of the largest available for faulty Stateflow models, as large datasets are uncommon in the context of CPSs. In addition, we evaluated three LLM variants, GPT-5.5, GPT-5.4-mini, and GPT-4.1-mini, to reduce the risk that the results reflect the behavior of a single model; however, the findings may not generalize to all LLMs or future model versions. Because the FlowRepair benchmark was publicly available before the evaluated models were used, training-data contamination cannot be excluded. Providing only component-level context avoids explicitly identifying the benchmark but does not rule out memorization of fault-fix patterns; successful repairs are therefore interpreted cautiously.

% Moreover, the potential presence of the FlowRepair benchmark in the LLMs' training data is partially mitigated by providing only the selected model component and its variable context, without benchmark or reference-patch information.

%\textbf{Internal Validity}: Another threat to the validity of our study is the parameter selection of our algorithms. In this work, we employ the same parameters as FlowRepair~\cite{arrieta2026flowrepair}, which were already tuned based on these same case studies. Although the one-hour budget enables an end-to-end comparison with FlowRepair, LLM inference latency may reduce the number of evaluated candidates; thus, our results reflect repair effectiveness under the same time budget rather than an equal number of candidate evaluations. As for the new LLM parameters, such as temperature, we used default parameters or based them on existing APR studies employing LLMs~\cite{Fan2023Automated}.

\textbf{Internal Validity}: Another threat concerns algorithm parameter
selection. We use the same parameters as FlowRepair~\cite{arrieta2026flowrepair},
which were tuned on the same case studies. The comparison evaluates end-to-end
repair effectiveness under the same one-hour wall-clock budget, but does not
isolate mutation quality from LLM inference latency because the approaches may
evaluate different numbers of candidates. LLM-specific parameters, such as
temperature, were set to default values or based on existing LLM-based APR
studies~\cite{Fan2023Automated}.

\textbf{Conclusion Validity}: Due to the stochastic nature of our approach, we repeated each run five times, following the same setup as FlowRepair~\cite{arrieta2026flowrepair}. With only five repetitions, the reported descriptive statistics provide a limited characterization of stochastic variability. We do not make inferential or statistical-significance claims; the comparisons are descriptive for this benchmark and configuration.

%\textbf{Construct Validity}: Our experimental evaluation is based on two metrics: Number of plausible patches generated and number of valid patches generated. These are the same metrics used by our baseline study FlowRepair~\cite{arrieta2026flowrepair}, and are commonly used in APR studies\cite{fan2023automated,jiang2021cure,lutellier2020coconut,xia2023automated}

\section{Related Work}
\label{sec:relatedwork}

Automated Program Repair (APR) has evolved from rule-based mutation techniques to learning-based and, more recently, Large Language Model (LLM)-driven approaches. Classical APR techniques primarily rely on search-based and constraint-based paradigms. Search-based methods generate candidate patches by applying predefined mutation operators and exploring the resulting search space guided by a correctness criterion, typically defined by a test suite~\cite{tan2018repairing,wen2018context,jiang2018shaping}. Foundational approaches such as GenProg leverage genetic programming to evolve program variants until a passing version is found~\cite{le2011genprog}. Constraint-based techniques, such as SemFix~\cite{nguyen2013semfix}, instead formulate repair as a constraint-solving problem derived from program executions. While effective, these approaches are limited by the scalability of the search space and their dependence on test suites and constraints, which restrict their applicability in complex or diverse scenarios~\cite{Fan2023Automated,xia2024automated}.

To address these limitations, learning-based approaches have been introduced, treating program repair as a data-driven task. Early work such as DeepFix learns transformations from historical bug-fix data~\cite{gupta2017deepfix}, and DeepRepair uses learned code similarity to guide repair ingredients \cite{White2019Sorting}, while Neural Machine Translation (NMT)-based approaches, such as SequenceR~\cite{8827954},  model repair as a translation problem from faulty to corrected code. These methods rely on datasets of previously fixed bugs to learn repair patterns~\cite{xia2024automated}. More recently, LLMs have enabled a new paradigm in which repairs are generated through prompting rather than task-specific training~\cite{brown2020language,prenner2022codex,Fan2023Automated}. Approaches such as AlphaRepair and ChatRepair further demonstrate the potential of LLMs for repair, including infilling-based fixes and iterative refinement using feedback~\cite{xia2022less,feng2020codebert,xia2024automated}. However, LLM-based repair approaches remain limited by issues such as sensitivity to prompt design, inconsistent outputs, and hallucinated fixes~\cite{10.1145/3571730}, and may not scale to the context of CPSs, as they require many patches to be validated through simulation-based testing.

In the context of CPSs, automated repair introduces additional challenges due to the tight coupling between software logic and physical processes, as well as the reliance on simulation-based validation. Prior work in CPS has largely focused on testing, including test generation and oracle design~\cite{7886937,menghi2020approximation}. More recently, FlowRepair demonstrated the feasibility of search-based APR for Simulink/Stateflow models using domain-specific mutation operators and simulation-driven objectives~\cite{arrieta2026flowrepair}. Despite the growing success of LLM-based repair in general-purpose software~\cite{10189263,Jiang2023Impact}, their application to CPS models remains limited due to domain-specific constraints and scalability challenges. Existing CPS repair approaches primarily target configuration or domain-specific issues rather than program-level repairs~\cite{10.1145/3395363.3397386,10.1145/3468264.3468601,10172628}, and only a limited number of works address repair in Simulink models~\cite{9211574,lyu2023autorepair}.

In this work, we build upon FlowRepair~\cite{arrieta2026flowrepair} and provide a controlled evaluation of LLM-based mutation for CPS repair, bridging the gap between mutation-based APR and LLM-driven approaches in this domain.

\section{Conclusion}
\label{sec:conclusion}
This work investigated the use of LLMs as mutation operators for APR in CPSs, within the FlowRepair framework for Stateflow models. We evaluated three LLM variants, GPT-5.5, GPT-5.4-mini, and GPT-4.1-mini, under the same repair setting used by FlowRepair. The results show that directly replacing several handcrafted mutation operators with LLM-based generation reduces repair performance in this setting. While the evaluated LLM variants can repair some localized semantic faults, such as expression simplifications and small arithmetic corrections, they were less effective for faults requiring precise value synthesis, missing expression insertion, structural changes, or consistent guided exploration. In our evaluated configuration, directly substituting FlowRepair's domain-specific operators with LLM-generated mutations led to less consistent repair outcomes. Effective integration may therefore require hybrid approaches that combine generative capabilities with domain constraints, structured operators, and feedback-driven optimization.\\

%This work investigated the use of LLMs as mutation operators for automated program repair in CPSs, within the FlowRepair framework for Stateflow models. Our results show that replacing handcrafted mutation operators with LLM-based generation degrades repair performance. While LLMs can handle localized semantic fixes, they struggle with precise edits, structural changes, and guided search. These findings suggest that LLMs are not a replacement for mutation operators in CPS repair. Instead, effective integration requires hybrid approaches that combine generative capabilities with domain constraints and feedback-driven optimization.

%\section*{Replication Package}

\noindent\textbf{Replication Package:} The replication package used is available on Zenodo:
\url{https://doi.org/10.5281/zenodo.21976165}

%The replication package used is available on Zenodo:
%\footnotesize{\url{https://zenodo.org/records/21976165}}

%For replicability, we provide the following package:\\ \footnotesize{\url{https://anonymous.4open.science/r/StateFlow-Repair-by-using-LLM-6FD7}}

\section*{Acknowledgment}
This work has been funded by the Spanish Ministry of Science, Innovation and Universities (project PID2023-152979OA-I00), funded by MCIU /AEI /10.13039/ 501100011033/FEDER, UE. The authors are part of the Software and Systems Engineering research group of Mondragon Unibertsitatea (IT1919-26), supported by the Department of Education, Universities and Research of the Basque Country.

\bibliographystyle{IEEEtran}
\bibliography{bibliography}

\end{document}